\documentclass[a4paper,11pt]{article}
\pdfoutput=1 

\usepackage{jcappub} 

\usepackage[T1]{fontenc} 
\usepackage{graphicx}
\usepackage{tikz}
\usetikzlibrary{calc,angles}
\usepackage{amsmath}
\usepackage{wrapfig}
\usepackage[utf8]{inputenc}

\newcommand{\dd}{\mathrm{d}}

\title{\boldmath Anisotropies of Ultra-high Energy Cosmic Rays Dominated by a Single Source in the Presence of Deflections}

\author{Andrej Dundovi\'c}
\author{and G\"unter Sigl}
\affiliation{II. Institut für Theoretische Physik, Universität Hamburg,\\
Luruper Chaussee, 149, 22761 Hamburg, Germany}

\emailAdd{andrej.dundovic@desy.de}
\emailAdd{guenter.sigl@desy.de}

\abstract{This work presents a scenario of ultra-high energy cosmic ray source distribution where a nearby source is solely responsible for the anisotropies in arrival directions of cosmic rays while the rest of the sources contribute only isotropically. An analytical approach focused on large-scale anisotropies, which are influenced by deflections in a Kolmogorov-type turbulent magnetic field, is employed to provide more general results. When the recent Pierre Auger Observatory angular power spectrum above 8 EeV is used the restricted model gives, under the assumption of the small angle approximation, a solution where the RMS deflection with respect to the line of sight is \(\alpha_{\rm rms} = \left(50^{+11}_{-10}\right)^\circ\), while the relative flux from the single source \(\eta=0.03\pm 0.01\). Furthermore, the solution can be translated into constraints on the source distance, luminosity, and extra-galactic magnetic field strength. For Centaurus A and the Virgo cluster the required relation between the coherence length and the RMS magnetic field strength is obtained: a coherence length of \(~\sim 100\,\mathrm{kpc}\) would imply the RMS field strength around \(1\,\mathrm{nG}\) for iron dominated and above \(10\,\mathrm{nG}\) for proton dominated composition. We also performed trajectory simulations with our publicly available code CRPropa to show that our analytical model can serve as a good approximation as long as the deflections in cosmic magnetic fields can be described as a random walk. The simulations showed that generally structured fields tend to suppress large-scale anisotropies, especially the dipole, compared to anisotropies at smaller scales described by higher multipoles.}

\begin{document}
\maketitle
\flushbottom

\section{Introduction}
\label{sec:intro}


A long-standing goal of the ultra-high energy cosmic ray research is to establish charged particle astronomy, that is, to use ultra-high energy cosmic rays (UHECRs) to study celestial objects and processes \cite{Sommers:2008ji,Kotera:2011cp}. The major known obstacle in this endeavour is the charged nature of cosmic rays which causes them to be deflected by cosmic magnetic fields. This fact minimises the possibility to directly pinpoint their sources with detected arrival directions \cite{Aharonian:2011da}. Without sufficient knowledge of the positions of UHECR sources and the properties of the intervening magnetic fields, it is a difficult, if not impossible, task to discriminate between the influence of the magnetic fields from the source distribution of UHECR by using the information on the arrival directions alone. Hopefully, constant advancements in complementary research areas, such as measurement of galactic and extra-galactic magnetic fields \cite{Widrow:2002ud, Pshirkov:2015tua}, gamma-ray astronomy \cite{Degrange:2015mda} and high-energy neutrinos \cite{Halzen:2016gng}, could eventually help to understand better the involved settings and processes which would, consequentially, lead to the identification of the sources.


It is self-explanatory that one of the most important observables in this research field is the arrival directions of UHECR. The arrival directions expressed in the terms of anisotropies, i.e., clustering of events in a particular region or part of the sky, could directly represent a signature of the sources. The experimental search for the anisotropies so far yielded results which indicate highly isotropic sky, both in the case of large-scale \cite{Aab:2014ila, ThePierreAuger:2014nja} and in the case of small-scale anisotropies \cite{Abreu:2010ab, PierreAuger:2014yba}. Only recently, the
Pierre Auger Observatory published the first statistically solid sign of the large-scale anisotropy in the arrival directions \cite{Aab:2017tyv} where the 6\% dipole anisotropy above 8 EeV pointing \(125^\circ\) away from the galactic centre is presented. Before it, there were indications of the dipole anisotropy reconstructed using three different techniques \cite{Aab:2016ban,ThePierreAuger:2014nja} and one joint analysis of Pierre Auger Observatory and Telescope Array above 10 EeV \cite{Deligny:2015vol} where the indication is verified with the full-sky coverage.


Taking into account the likely premise that galactic magnetic fields are not strong enough to be the cause of the isotropy itself \cite{Giacinti:2010dk,Jansson:2012pc}, reasons for such remarkable isotropy could be found either in strong extra-galactic deflections of UHECRs or homogeneously distributed low-luminous sources or both combined. The majority of the acceleration models rely on already known, electromagnetically visible, objects which are following the matter distribution of the local universe. Since the local universe is not quite homogeneous \cite{Dolag:2004kp}, the assumption of homogeneously distributed sources, in that case, is unrealistic. The already cited lack of correlation of known catalogue sources or matter distribution with the observed arrival directions could be a sign of relatively strong deflections which are caused by either a heavier composition, as also supported by experimental data \cite{Aab:2016htd}, or considerable magnetic field strength, or both.


From the perspective of extra-galactic magnetic field models, deflections of UHECRs range from weak \cite{Dolag:2003ra} to strong \cite{Sigl:2003ay}. The newest ESA's Planck satellite measurements of primordial magnetic fields imply upper limits of nano gauss levels in voids \cite{Ade:2015cva} which constrain those models. Examples of new models which are in accordance with those limits are \cite{Hackstein:2016pwa,AlvesBatista:2017vob}. The magnetic fields play a complex role in the anisotropy formation, and this role is still poorly understood, but several characteristics are well established. For example, one is the claim that magnetic fields, by themselves, cannot create anisotropies from an underlying isotropic distribution due to Liouville's theorem \cite{9780521016469}, but they can generate small-scale anisotropies from pre-existing large-scale dipolar anisotropies \cite{Giacinti:2011mz}.


To find the answer what the realistic deflection is, one could take the fact that the observed isotropy is relatively high and that no obvious model can explain it as a consequence of the surrounding universe, i.e., our galaxy and its neighbourhood of several megaparsecs. As a side note, an exotic acceleration mechanism of homogeneously distributed accelerators within the galaxy or its galactic halo could still explain the observables, but the base argument of this paper is not focused on that possibility following reasoning of many other papers which also do not consider that sources are homogeneously distributed within the galaxy or in the surrounding universe (see, for example, \cite{berezinsky2008propagation,Bhattacharjee:1998qc,hillas1984origin,Nagano:2000ve}). For a recent review about the galactic-source conjecture see \cite{YounkICRC2013}.


The assumption of inhomogeneously distributed sources in the surrounding universe which follows the visible matter\footnote{A possible, but extreme case could be a source nearby which is ``invisible'', like an active galactic nucleus which was active \(10^8\) - \(10^9\) years ago and cannot be detected in photons anymore, but cosmic rays from it are reaching Earth just now as they are delayed compared to photons \cite{Berezinskii:1990ap}.} statistically implies that some of the sources or the groups of sources are closer to our galaxy than the others. Therefore, the total UHECR flux may be dominated by sources nearby whose flux is the least attenuated by interactions. Given the observed isotropy, one could argue if anything has a chance of producing an anisotropic pattern, such as a dipolar anisotropy, the source nearby would be one of the most viable options. Firstly, far away sources produce a mostly isotropic flux since there are many of them and their distribution approaches homogeneity at large scales, and secondly, due to longer paths through intervening extra-galactic magnetic fields with a turbulent component, more remote sources tend to spread their arrival directions widely. Full-scale Monte Carlo simulations also support this conclusion \cite{Hackstein:2016pwa}, even with relatively weak magnetic fields in voids compared to other papers. Thus, the question is whether sources nearby such as Centaurus A (Cen A) or the Virgo cluster can cause anisotropies and what would be their shape. It is worth mentioning that the scenario with a source nearby, or particularly, Centaurus A and the Virgo cluster, has been studied many times before with different approaches and various goals: in diffusion models of propagation \cite{Berezinskii:1990ap, Harari:2013pea,Harari:2015hba}, by simulating ballistic propagation \cite{Isola:2001ng,Dolag:2008py,Kim:2012rp,Fargion:2008sp}, in the context of galactic magnetic field models \cite{Farrar:2012gm,Keivani:2014kua}, by constraining their flux to the observational data \cite{Abreu:2010ab,Kachelriess:2008qx}, etc. A more general constraint of a single source model independent of the specific position, luminosity, and magnetic field strength or structure is so far missing. With this work, we try to fill this gap by focusing solely on the anisotropic pattern left by a source nearby and relying mainly on analytical tools.

The present work is developed around the scenario in which a source nearby contributes a significant fraction of the total UHECR flux while the rest of sources contribute isotropically. The focus here is on the large-scale anisotropies since they can be experimentally better determined due to a high number of events which they include and they are the least sensitive to the coherent magnetic deflections which are poorly constrained. The angular power spectrum, often used as a measure of anisotropy, is the primary method of quantifying anisotropies here because it is rotationally invariant, thus enabling a more general analysis without focusing on specific directions on the sky. Together with analytical expressions, this approach gives not only real constraints on the scenarios with a source nearby but also a methodological guide for a more sophisticated Monte Carlo approach which is often the only viable analysis technique in the field.

The organisation of the paper is the following: in section \ref{sec:model} a simple model of a single source with an isotropic background is presented, followed by section \ref{sec:parametrization} where the parametrisation of the model is considered. In section \ref{sec:montecarlo} comparison with Monte Carlo methods is performed to show the robustness of the introduced simplifications and to analyse how structured magnetic fields can change the picture, in section \ref{sec:constraints} the model is applied to specific sources to constrain physical parameters based on the experimental angular power spectrum data. The paper ends with section \ref{sec:conclusions} -- conclusions.

\section{Analytical model of a single source}
\label{sec:model}

As already mentioned in the introductory part, the high-level of isotropy of the observed sky could be caused by extra-galactic magnetic fields strong enough to wash out the memory of the UHECR origin. The heavier composition at the end of the spectrum \cite{PhysRevD.90.122006} makes this interpretation more likely as it increases the average deflections. Motivated by these arguments, we consider the possibility that a source nearby could cause the observed large-scale anisotropy whose image is broadened by deflection in the magnetic fields.

Here we take into account only the omnipresent turbulent magnetic field component and neglect any regular components that depend on a specific structural magnetic field model to keep the analysis less model dependent. For the description of propagation through a turbulent magnetic field, we rely on the results of \cite{Achterberg:1999vr} in which cosmic ray propagation in a random field is described by using a stochastic differential equation of the It\^{o} form (see textbook \cite{GardinerStochastic2009}):
\begin{equation}
\label{eq:sde}
 \dd n_i = -2\mathcal{D}_0 n_i \dd s + \sqrt{2\mathcal{D}_0} P_{ij}(\mathbf{\hat{n}})\dd W_j
\end{equation}
in which \(\mathbf{\hat{n}} = (n_1, n_2, n_3)\) is the unit vector along the flight direction, \(\dd s \equiv c\dd t\) is the differential path, \(P_{ij}(\mathbf{\hat{n}}) \equiv \delta_{ij} - n_i n_j\) is the projection tensor onto the plane perpendicular to \(\mathbf{\hat{n}}\), and \(\dd \mathbf{W} \equiv (\dd W_1, \dd W_2, \dd W_3)\) is a Wiener process satisfying \(\langle \dd W_i\rangle = 0\), \(\langle \dd W_i \dd W_j\rangle = \delta_{ij} \dd s\). The quantity \(\mathcal{D}_0\) represents a scalar diffusion coefficient in the flight direction.

When a cosmic ray traverses a turbulent magnetic field, and if scales of the turbulence roughly match the cosmic ray's gyroradius, it erratically changes its direction of flight \(\mathbf{\hat{n}}\), and hence performs Brownian motion on the unit sphere of its flight direction starting from the initial direction \(\mathbf{\hat{n}}_0\) \cite{LongairHigh2011,ChandrasekharStochastic1943}. The Brownian motion distribution is the one that describes a distribution of ensemble-averaged flight directions of particles that propagated from the same initial direction \cite{mardia2000directional,FrancisEtude1928}.

In the context of distributions on the unit sphere, one can define the so-called mean resultant length as a measure of the angular spread of events on a sphere defined as \(\rho = \frac{1}{N}||\sum_{i=1}^N \mathbf{\hat{n}}_i ||\) where \(\mathbf{\hat{n}}_i\) are unit vectors representing \(N\) events. In case when the spread of events is distributed around the direction \( \mathbf{\hat{r}} \), the mean resultant length can be expressed through the tangent-normal decomposition of the unit vectors \cite[ch. 9]{mardia2000directional}:
\begin{align}
\label{eq:meanresultantlength}
  \rho = \frac{1}{N}\left|\left|\sum_{i=1}^N \mathbf{n}_i \right|\right| = \frac{1}{N} \left|\left| \sum_{i=1}^N t_i \mathbf{\hat{r}} + (1-t_i^2)^{\frac{1}{2}} \mathbf{\hat{r}}_\bot \right|\right| = \frac{1}{N} \sum_{i=1}^N t_i = \langle \mathbf{\hat{n}}\cdot \mathbf{\hat{r}} \rangle
\end{align}
where \(t_i = \mathbf{\hat{n}}_i \cdot \mathbf{\hat{r}}\). The terms next to \(\mathbf{\hat{r}}_\bot\) vanish after averaging as the distribution is invariant under rotation around \(\mathbf{\hat{r}}\) . The behaviour of the previous formula agrees with intuitive reasoning that the spread of directions on a sphere after particles are propagated for a sufficient time leads to the loss of correlation \(\langle \mathbf{\hat{n}}\cdot \mathbf{\hat{r}} \rangle = 0 \Rightarrow \rho = 0\), while on the other hand, without any deflections all events are correlated with the line of sight making \(\rho = 1\).

To relate eq. (\ref{eq:meanresultantlength}) and the physical process of random walk in flight directions represented by eq. (\ref{eq:sde}), we use results from \cite[Appendix B]{Achterberg:1999vr}. An arrival direction \(\mathbf{\hat{n}}_i\) and the unit vector \(\mathbf{\hat{r}}\) that points along the line of sight from the source to the observer define an angle \(\alpha\): \(\cos \alpha_i \equiv \mathbf{\hat{n}}_i \cdot \mathbf{\hat{r}}\). It is shown in \cite{Achterberg:1999vr} that generally, \(\langle \mathbf{\hat{n}} \cdot \mathbf{\hat{r}} \rangle\) is a function of \(\mathcal{D}_0 L\) where \(L\) is the distance between the source and the observer. Specifically, under the small angle approximation the root mean square value of the angle \(\alpha\) reads \(\alpha^2_{\rm rms} = \frac{4}{3}\mathcal{D}_0 L\) and that will be used later in this work. The small angle approximation in this case is justified up to a few \(\mathcal{D}_0 L\) as it will be shown later numerically (the right panel of fig. \ref{fig:LD0}).


It is proved in \cite{RobertsRandom1960} that the Brownian motion distribution can be approximated by a Fisher distribution (or Fisher -- von Mises distribution on the \((p-1)\)-dimensional sphere where \(p=3\)) if they share the same mean resultant length \(\rho\). A Fisher distribution centred around \(\hat{\mathbf{r}}\) is defined by \cite{fisher1953dispersion}:
\begin{equation}
f_{\rm FvM}(\hat{\mathbf{n}}| \kappa) = \frac{\kappa}{4\pi \sinh(\kappa)} \exp(\kappa \hat{\mathbf{n}}\cdot\hat{\mathbf{r}})
\label{eq:fisher}
\end{equation}
where \(\kappa\) is the concentration parameter (later also called the spread parameter) determining the spread of events. Physically, $\kappa$ encodes everything that influences the angular spread, in particular, the distance from the source combined with magnetic fields, that is, $\kappa$ equals zero in the limit of extremely strong magnetic fields which totally erase the memory of the source position, and the distribution becomes just a uniform distribution. In the absence of magnetic fields, $\kappa \gg 1$, the distribution behaves like a point source.

The mean resultant length for a Fisher distribution is defined by the following equation:
\begin{equation}
\label{eq:kapparho}
 \rho \equiv A_3(\kappa) = \coth(\kappa) - \frac{1}{\kappa} \ .
\end{equation}
Its two important limiting cases are \(A_3(\kappa) \approx \kappa/3\) for \(\kappa \ll 1\) and \(A_3(\kappa) \approx 1 - 1/\kappa\) for \(\kappa \gtrsim 1\) \cite[Appendix 1]{mardia2000directional}. Now, since the two mentioned distributions share the same resultant length, it follows that the concentration parameter is determined directly by \(\kappa = A^{-1}_3(\langle \mathbf{\hat{n}}\cdot \mathbf{\hat{r}} \rangle)\).

In this work, trajectory simulations are performed in which particles are propagated through various instances of turbulent magnetic fields to test the validity of this distribution. There are no signs that the distribution of deflection angles of propagated particles is violating a Fisher distribution for any relevant set of physical parameters as long as the particles cross many coherence lengths of a turbulent magnetic field (the left panel of figure \ref{fig:LD0}). Further details about numerical checks against the simulations are presented in sec. \ref{sec:parametrization}.

The reader's attention is drawn to the fact that the expression (\ref{eq:meanresultantlength}) and the given distribution, eq. (\ref{eq:fisher}), when transformed to the perspective of a fixed observer on some distance from a source, represent the source which is injecting particles in a single (beamed) direction along the line of sight from the source to the observer. If the source injects particles in other directions in addition to the direction along the line of sight, those particles could also contribute to the distribution of arrival directions and which generally differs from eq. (\ref{eq:fisher}). The new distribution can be constructed by directly integrating contributions from those other directions:
\begin{equation}
\label{eq:general_distribution}
 f(\hat{\mathbf{n}}| \kappa) = \mathcal{N}(\kappa) \int_\Omega \exp(\kappa \hat{\mathbf{n}}\cdot\hat{\mathbf{r'}}) p\left(\mathbf{r}' | \kappa \right) \dd \mathbf{r}'
\end{equation}
where \(\mathcal{N}(\kappa)\) is the norm while \(p\left(\mathbf{r}' | \kappa \right)\) is a probability distribution for directions \(\mathbf{r}'\) which depends also on the concentration parameter \(\kappa\). For example, in case of minor deflections the distribution tends to \(p\left(\mathbf{r}' | \kappa \right) \propto \delta(\mathbf{r}' - \mathbf{r})\), i.e., to eq. (\ref{eq:fisher}). Oppositely, for significant spreads, \(\kappa \ll 1\), the expression tends to the uniform distribution as same as the original Fisher distribution.

\begin{figure}
\begin{minipage}{0.50\textwidth}
	      \centering
	      \includegraphics[scale=0.40]{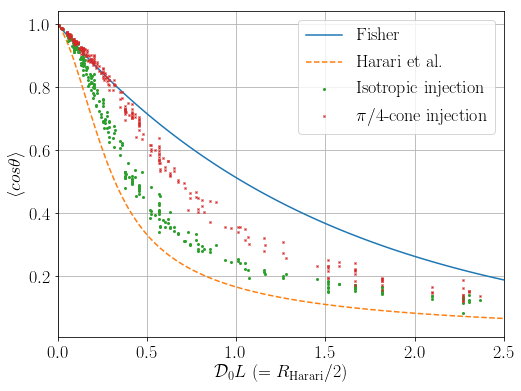}	  
\end{minipage}
\begin{minipage}{0.50\textwidth}
	      \centering
	      \includegraphics[scale=0.40]{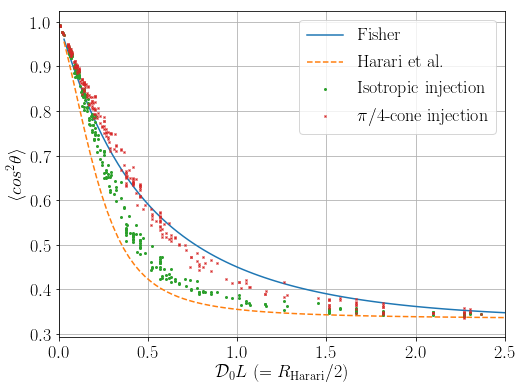}
\end{minipage}
\caption{These two panels, in analogy to fig. 2 in \cite{Harari:2015mal}, represent average \(\cos\theta\) (the left panel) and \(\cos^2\theta\) (the right panel) calculated from four distinct distributions as a function of \(\mathcal{D}_0 L\). The blue line is from the Fisher distribution, eq. (\ref{eq:fisher}), the orange one is the distribution assumed in \cite{Harari:2015mal}. The green and red dots are obtained from events of a simulation in which the single source injects particles isotropically and beamed within a \(\pi/4\)-cone directed towards the observer. The narrowed the cone is the better the distribution is described by eq. (\ref{eq:fisher}).\label{fig:comparison_with_harari}}
\end{figure}

A similar approach to modelling the arrival direction distribution of a single source is given in \cite{Harari:2015mal} with the difference that the authors of that paper constructed a distribution without giving a prior physical rationale, only by providing the convincing fits to their simulations. In their case, the distribution is a Fisher distribution, eq. (\ref{eq:fisher}), combined with the isotropic term which is ``mostly from particles (from a single source) that diffused long times and made several turns before reaching the observer'' (quoted from \cite{Harari:2015mal}). One could obtain the isotropic term with analogues properties from eq. (\ref{eq:general_distribution}) by expanding \(p\left(\mathbf{r}' | \kappa \right) \approx 1 + \kappa \cdot \partial p / \partial \kappa\) for small \(\kappa\). Integrating the first term would give the isotropic term which depends only on \(\kappa\). Here, we will not follow the same path of modelling due to two reasons. One is that the constructed distribution from \cite{Harari:2015mal} is departing from numerically obtained results from our simulations, especially for \(\mathcal{D}_0 L \le 1\) (which is labeled in \cite{Harari:2015mal} as the normalised source distance, \(R_{\rm Harari} = 2 \mathcal{D}_0 L\)) as can be seen in fig. \ref{fig:comparison_with_harari}. The other, more important reason, is that by fitting parameters of the distribution to the numerical results we can no longer directly relate, using eq. (\ref{eq:kapparho}), the concentration parameter \(\kappa\) to physical quantities that are the cause of the spread in arrival directions.

From now on, we consider only a single source which can be accurately modelled by eq. (\ref{eq:fisher}), i.e., which is either beamed towards the observer or which image is not broadened by deflections significantly. By doing that we still preserve the analytical transparency of the results. More general considerations will be left for future works.

If one takes into account other, more distant sources which contribute only isotropically to the arrival directions, an image of a single (beamed) source nearby described with a Fisher distribution is expanded with a normalised constant isotropic term. Hence, the total distribution reads:
\begin{equation}
f_\mathrm{src+bg}(\hat{\mathbf{n}}|\eta, \kappa) = \frac{1-\eta}{4\pi} + \eta \frac{\kappa}{4\pi \sinh(\kappa)} \exp(\kappa \hat{\mathbf{n}}\cdot\hat{\mathbf{r}}_\mathrm{src})
\label{eq:distribution}
\end{equation}
where $\eta$ and $\mathbf{r}_\mathrm{src}$ are, respectively, the relative flux  $j_\mathrm{src}/j_\mathrm{tot}$ and the direction of the source nearby.

\subsection{The angular power spectrum}

Following an analytic approach while keeping the single source location unspecified, we choose the angular power spectrum as the main observable for the model. The angular power spectrum $C_\ell$ is defined through the expansion of the distribution function in spherical harmonics:
\begin{equation}
\label{eq:ftosharmonics}
 f(\theta,\varphi)=\sum_{\ell=0}^\infty \sum_{m=-\ell}^\ell a_{\ell m} \, Y_\ell^m(\theta,\varphi)
\end{equation}
\begin{equation}
\label{eq:powerspectrum}
 C_\ell \equiv \langle |a_{\ell m}|^2 \rangle = \frac{1}{2\ell +1}\sum_{m=-\ell}^\ell | a_{\ell m} |^2
\end{equation}
where $f(\theta,\varphi)$ is a distribution of the arrival directions, $Y_\ell^m$ are spherical harmonics and $a_{\ell m}$ coefficients defined as $a_{\ell m}=\int_{\Omega} f(\theta,\varphi)\, Y_\ell^{m*}(\theta,\varphi)\,d\Omega$.

\begin{figure}
\begin{minipage}{0.50\textwidth}
	      \centering
	      \includegraphics[scale=0.40]{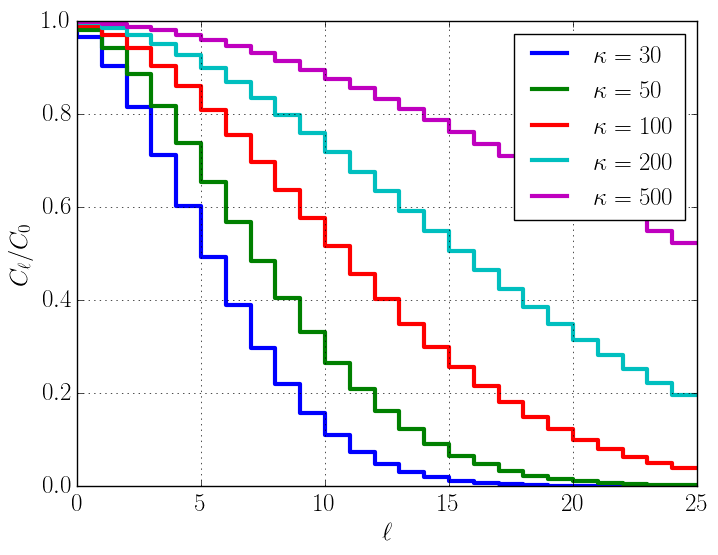}	  
\end{minipage}
\begin{minipage}{0.50\textwidth}
	      \centering
	      \includegraphics[scale=0.40]{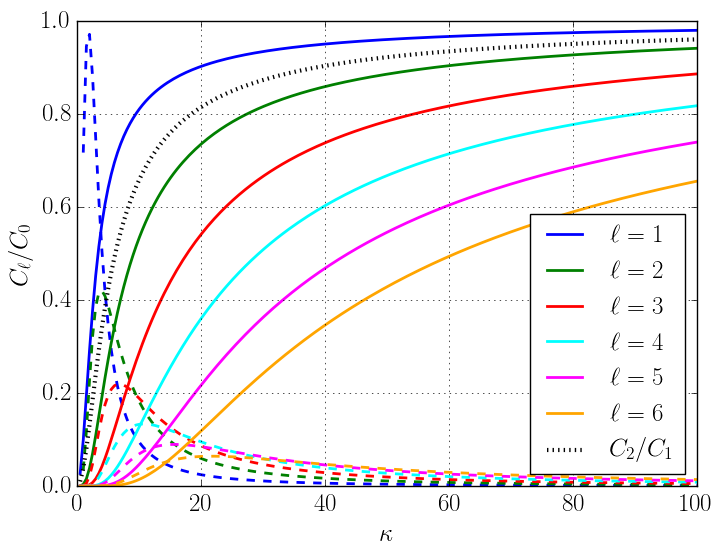}
\end{minipage}
\caption{These two panels illustrate solutions of the single source model (\ref{eq:vonmisses_solution}) without the background contribution (\(\eta=1\)). The first one shows the dependence of the normalized power spectrum \(C_\ell\) on the multipole moment \(\ell\) for different parameter \(\kappa\), while the second one shows the dependence of the power spectrum on $\kappa$ for first few moments; the dashed curves are derivatives of solid lines multiplied by factor 10; the black dotted line represents the ratio \(x=C_2/C_1\) versus \(\kappa\). All figures in this work are plotted using \cite{Hunter:2007}.\label{fig:vonmisses_plot}}
\end{figure}

Since the angular power spectrum is rotationally invariant, without loss of generality the observer can be placed at the origin of the coordinate system and the \(z\)-axis can be chosen along the source direction: $\hat{\mathbf{n}}\cdot\hat{\mathbf{r}}_\mathrm{src} = \cos(\theta)$. Inserting (\ref{eq:distribution}) in (\ref{eq:powerspectrum}) gives:
\begin{align}
 a_{\ell m} =&  \frac{1-\eta}{4\pi}\int_0^{2\pi}d\varphi\int_0^\pi \,d\theta\,\sin\theta Y_\ell^{m*} (\theta,\varphi) \nonumber \\
	    & + \eta\frac{\kappa}{4\pi \sinh(\kappa)} \int_0^{2\pi}d\varphi\int_0^\pi \,d\theta\,\sin\theta \exp(\kappa \cos(\theta)) Y_\ell^{m*} (\theta,\varphi) \\
	    =& \delta_{\ell 0}\frac{1-\eta}{4\pi}2\sqrt{\pi} + \delta_{m 0} \sqrt{\frac{2\ell+1}{4\pi}} \int_0^\pi \,d\theta\,\frac{\kappa \sin(\theta)}{2 \sinh(\kappa)} \exp(\kappa \cos(\theta)) \, P_\ell ( \cos{\theta} )
\end{align}
The last integral \(\mathcal{I}_\ell =  \frac{\kappa}{2 \sinh(\kappa)}\int_{-1}^1 du \exp(\kappa u) P_\ell(u)\) can be reduced to a recurrence relation using the property of Legendre polynomials $(2n+1) P_n(u) = \frac{d}{du} \left[ P_{n+1}(u) - P_{n-1}(u) \right]$:
\begin{align}
\label{eq:recurrence}
 \mathcal{I}_0 =& 1\ , \quad \mathcal{I}_1 = \coth{\kappa} -\frac{1}{\kappa}\ , \quad \mathcal{I}_{\ell} =  \mathcal{I}_{\ell-2} -\frac{2\ell-1}{\kappa} \mathcal{I}_{\ell-1}\ .
\end{align}
The recurrence yields the final form expressed in terms of the modified Bessel function of the first kind:
\begin{align}
\mathcal{I}_\ell =& \sqrt{\frac{\pi}{2}} \frac{\sqrt{\kappa}}{\sinh{\kappa}} I_{\ell+\frac{1}{2}}(\kappa) \\
a_{\ell 0} =& \delta_{\ell 0}\frac{1-\eta}{\sqrt{4\pi}} + \eta\sqrt{\frac{2\ell+1}{4\pi}} \sqrt{\frac{\pi}{2}} \frac{\sqrt{\kappa}}{\sinh(\kappa)} I_{\ell+\frac{1}{2}}(\kappa)
\label{eq:a_l0_solution}
\end{align}
The final power spectrum reads
\begin{align}
\label{eq:vonmisses_solution}
 C_{\ell} =&\begin{cases}
	     \frac{1}{4\pi} &\text{if } \ell=0 \\ 
             \eta^2 \frac{1}{4\pi}\mathcal{I}_\ell^2 = \eta^2 \frac{\kappa}{8}\sinh^{-2}(\kappa) I^2_{\ell+\frac{1}{2}}(\kappa) &\text{if } \ell > 0
            \end{cases}
\end{align}
where it is advisable for numerical calculations to use the recurrence form (\ref{eq:recurrence}) instead of the last form due to numerical instability around \(\kappa \approx 0\).

The solution for \(\eta=1\) is plotted in figure \ref{fig:vonmisses_plot} for the first few moments and for several parameters $\kappa$. The figure shows consistency with limiting cases: when the angular spread is high enough, or the spread parameter is low enough, to erase the position of the source completely, only the monopole component will be present (\(\forall \ell > 0 \quad \lim_{\kappa\rightarrow 0} C_\ell = 0\)). In the opposite limit, without any deflections, the source will be in the form of a delta function and all moments will be equal. The parameter $\eta$ mainly governs the ratio of higher multipoles to the monopole.

\begin{figure}
\begin{minipage}{0.50\textwidth}
	      \centering
	      \includegraphics[scale=0.40]{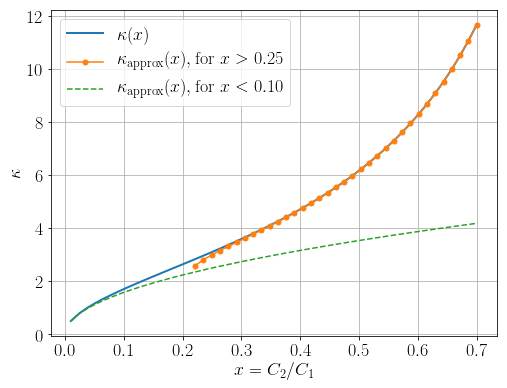}
\end{minipage}
\begin{minipage}{0.50\textwidth}
	      \centering
	      \includegraphics[scale=0.40]{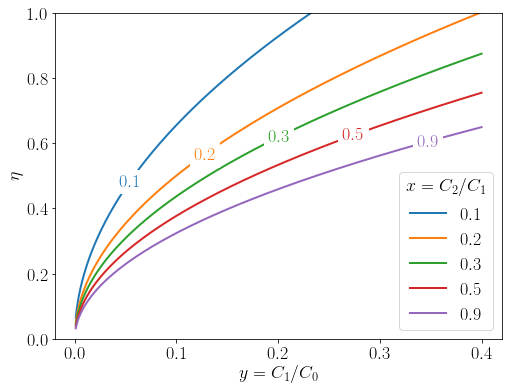}
\end{minipage}
\caption{The left panel shows how to retrieve parameter \(\kappa\) from the ratio of the quadrupole and the dipole (\(x=C_2/C_1\)) with the two analytical approximations described in the footnote. The right panel shows how \(\eta\), which parametrises the flux contribution of the single source, can be determined from \(x=C_2/C_1\) and the relative dipole amplitude \(y=C_1/C_0\).\label{fig:inverse_solution}}
\end{figure}

The parameters of the model, \(\kappa\) and \(\eta\), can be determined from the ratios of multipoles. In particular, the ratio of the quadrupole and the dipole, \(x=C_2/C_1\), is given by:
\begin{equation}
\label{eq:mukapa}
 \frac{C_2}{C_1} \equiv x = \left[\left(\coth{\kappa }-\frac{1}{\kappa}\right)^{-1} - \frac{3}{\kappa}\right]^2
\end{equation}
In figure \ref{fig:inverse_solution} we solve eq. (\ref{eq:mukapa}) for \(\kappa\) numerically\footnote{Nevertheless, useful analytic expressions in case of low (\(x < 0.1\)) and high (\(x > 0.25\)) ratio limits can be obtained:
\begin{equation}
 \kappa(x) \approx \kappa_{\rm approx}(x) = \begin{cases}
	    \frac{3-\sqrt{x} + \sqrt{x+6\sqrt{x}-3}}{2-2\sqrt{x}}, & \text{valid for } x \gtrsim 0.25, \\ \nonumber
            5 \sqrt{x}, & \text{valid for } x \lesssim 0.1\,.
          \end{cases}
\end{equation}
}. The derivative of \({\rm d}x(\kappa)/{\rm d}\kappa\) is always positive which confirms the monotonous behaviour of the function and implies the existence of a unique solution for \(\kappa\) in the domain of interest. The parameter \(\eta\) is then expressed through \(y=C_1/C_0\) and \(\kappa(x)\) (figure \ref{fig:inverse_solution}):
\begin{equation}
\label{eq:mukapa2}
 \eta\left(y, x\right) = \frac{\sqrt{y}\kappa(x)}{|\kappa(x) \coth{\kappa(x)}-1|}\ .
\end{equation}
Our model gives \(y = C_1/C_0 \leq 1\) which implies \(\eta \leq \left[\coth(\kappa)-1/\kappa\right]^{-1}\). On the right panel of figure \ref{fig:inverse_solution} this inequality is violated for certain pairs \(x\) and \(y\) where \(\eta\) becomes larger than unity which is unphysical since it would result in negative flux in certain directions, c.f. eq. (\ref{eq:distribution}).

\section{Interpreting the spread parameter in terms of physical quantities}
\label{sec:parametrization}

The blurring of source images is caused by deflections of cosmic rays in intervening magnetic fields. Interactions can also contribute to the angular spread, but for highly energetic particles their effect can be neglected as the scattering angle is of the order of the inverse Lorentz factor and can thus be neglected. Relevant quantities for deflections in magnetic fields are the cosmic ray rigidity $\mathcal{R}=E/Ze$ and the properties of the turbulent magnetic field.

In the small angle approximation, from eq. (\ref{eq:sde}), as shown in \cite[see eq. B21 and B22]{Achterberg:1999vr}, and eq. (\ref{eq:kapparho}) follows:
\begin{equation}
\label{eq:smallangle}
 \langle \mathbf{\hat{n}}\cdot \mathbf{\hat{r}} \rangle \approx 1 - \frac{2}{3}\mathcal{D}_0 L \Rightarrow \kappa \approx \frac{3}{2}(\mathcal{D}_0 L)^{-1} \ .
\end{equation}

If the turbulent field is given as a power law, like a Kolmogorov type of turbulence, the scalar diffusion coefficient of the flight direction in the small angle approximation yields \cite{Achterberg:1999vr}:
\begin{equation}
\label{eq:D0_physical}
 \mathcal{D}_0 = \frac{1}{8}\left(\frac{c}{\mathcal{R}}\right)^2 B^2_{\rm rms} L_c
\end{equation}
where \(B_{\rm rms}\) is the root mean square of magnetic field strength and \(L_c\) is the field's coherence length. From this, one can show that the root mean square deflection angle
in the flight direction of UHECR after traversing a distance \(L\) is described with:
\begin{equation}
 \theta_{\rm rms}^2 \approx 4\mathcal{D}_0 L = \frac{1}{2}\left(\frac{c}{\mathcal{R}}\right)^2 B^2_{\rm rms} L_c L
\end{equation}
or
\begin{equation}
 \alpha_{\rm rms}^2 \approx \frac{4}{3}\mathcal{D}_0 L = \frac{1}{6}\left(\frac{c}{\mathcal{R}}\right)^2 B^2_{\rm rms} L_c L
\end{equation}
when considering the deflection in respect to the line of sight from the source to the observer. Besides \cite{Achterberg:1999vr}, these expressions are derived also in \cite{astro-ph/0202362v2}.

The spread parameter \(\kappa\) is then equated with:
\begin{align}
\label{eq:parametrization}
 \kappa \approx 12\mathcal{R}^2 B_{\rm rms}^{-2}(L_c L)^{-1} &= 2\alpha_{\rm rms}^{-2} \nonumber \\
  &= 140.2 \cdot \left(\frac{\mathcal{R}}{\mathrm{EV}}\right)^2 \left(\frac{B_\mathrm{rms}}{\mathrm{nG}}\right)^{-2} \left(\frac{L}{\mathrm{Mpc}}\frac{L_c}{100\mathrm{kpc}}\right)^{-1}\ .
\end{align}

Cosmic rays that are deflected in one realization of a turbulent field will not generally be consistent with the smeared shape of the Fisher distribution if they are not sufficiently randomised by traversing multiple coherence lengths of the turbulent magnetic field defined as in \cite{astro-ph/0202362v2}:

\begin{equation}
 L_c = \frac{1}{2} L_\mathrm{max}\cdot \frac{\alpha-1}{\alpha}\cdot \frac{1-\left(\frac{L_\mathrm{min}}{L_\mathrm{max}}\right)^\alpha}{1-\left(\frac{L_\mathrm{min}}{L_\mathrm{max}}\right)^{\alpha-1}}\ .
\end{equation}
where the spectral index \(\alpha = 5/3\) for the Kolmogorov spectrum and \(L_{\rm min}\) and \(L_{\rm max}\) are the minimal and maximal length scales of the magnetic field power spectrum.

Now it remains to check ranges of the physical parameters for which the small angle approximation holds and also to check the general consistency of these analytical expressions against Monte Carlo simulations. We employ a publicly available Monte Carlo framework for the propagation of cosmic rays CRPropa version 3 \cite{batista2016crpropa} for every numerical calculation that follows. In CRPropa, charged particles are propagated by solving the equations of motion of a relativistic charged particle in a magnetic field. The solving algorithm is based on the Runge-Kutta class of methods, in particular, the Cash-Karp method. A turbulent magnetic field is generated first in the k-space by placing random amplitudes on a grid following a given power spectrum law. The grid is then modulated by a random complex phase, and a random orientation with \(\mathbf{k}\cdot\mathbf{B} = 0\) to satisfy \(\nabla \cdot \mathbf{B} = 0\). In the end, the field is transformed into real space with a fast Fourier transform and normalised to \(B_{\rm rms}\).

\begin{figure}
\begin{minipage}{0.50\textwidth}
	      \centering
	      \includegraphics[scale=0.45]{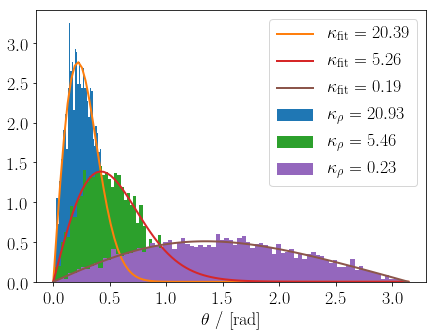}
\end{minipage}
\begin{minipage}{0.50\textwidth}
	      \centering
	      \includegraphics[scale=0.45]{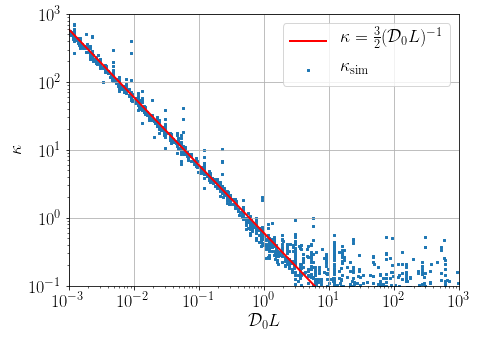}	  
\end{minipage}
\caption{The left panel shows the binned deflection angles of three simulations where each has a different set of physical parameters that determine the particle propagation. The solid lines are plotted Fisher distributions in which \(\kappa\) is determined from eq. (\ref{eq:meanresultantlength}) and eq. (\ref{eq:kapparho}). The right panel displays the dependence of \(\kappa\) on \(\mathcal{D}_0 L\) where the blue dots represents simulations in which \(\kappa\) and \(\mathcal{D}_0 L\) are determined by eq. (\ref{eq:kapparho}) and (eq. \ref{eq:D0_physical}), respectively. The red line is a plot of eq. (\ref{eq:smallangle}) which shows the validity of the small angle approximation.\label{fig:LD0}}
\end{figure}

The simulations are performed on a range of parameters by injecting \(2000\) protons in each simulation. The particles are propagated radially from a centre of a sphere and detected whenever crossing the border of the sphere, from the inside or the outside, and discarded once they reached the maximum trajectory length of 20 Mpc. The parameters range from \(\{0.001, \dots, 100\}{\rm EeV}\) in the energy of particles, \(\{0.1, \dots, 5000\}{\rm nG}\) in the magnetic field strength, \(\{2, \dots, 30\}{\rm kpc}\) in the coherence length, and \(\{10, \dots, 4000\}{\rm kpc}\) in the radius of the sphere. The spread parameter is calculated from the mean resultant length of detected flight directions by using eq. (\ref{eq:kapparho}). The left panel in fig. \ref{fig:LD0} shows distributions of the deflection angle \(\alpha\) for three simulations to show how well the Fisher distribution fits the Monte Carlo generated data. The spread parameters which are displayed in the legend correspond to the ones calculated from eq. (\ref{eq:kapparho}) (\(\kappa_\rho\)), and fitted on the Fisher distribution (\(\kappa_{\rm fit}\)). The right panel shows that the small angle approximation in eq. (\ref{eq:smallangle}) well describes the spread parameter \(\kappa\) for \(\mathcal{D}_0 L \lesssim 1\). Entering in a highly diffusive regime where \(\mathcal{D}_0 \gg 1\), ballistic propagation in the numerical code becomes highly inefficient and so the number of recorded events drops. The reduced number of recorded events increase statistical fluctuations. 

\begin{figure}
 \centering
 \includegraphics[scale=0.50]{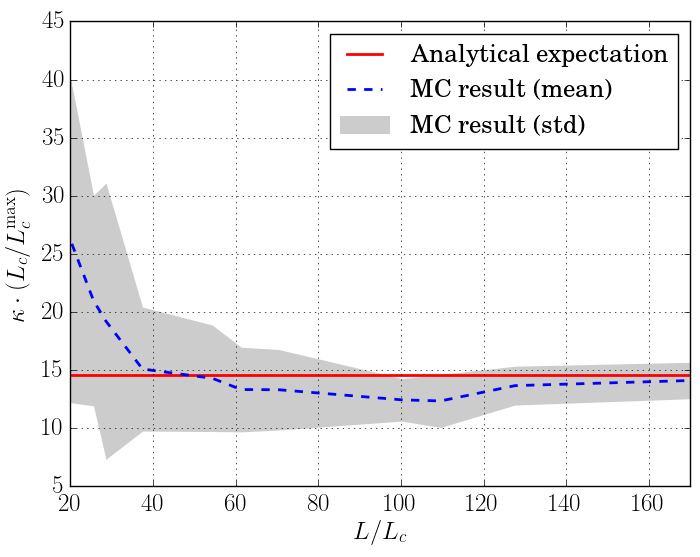}
 \caption{\label{fig:coherencelengths} CRPropa version 3 \cite{batista2016crpropa} is employed to check numerically how many coherence lengths cosmic rays needs to traverse to be sufficiently randomised to match the predicted value from the Fisher distribution. The analytical expectation follows from equation (\ref{eq:parametrization}) where the result is multiplied by \(L_c/L_c^\mathrm{max}\) to remove the dependence of \(\kappa\) on \(L_c\) to stress the discrepancy; \(L_c^\mathrm{max} = 256 \mathrm{kpc}\) but it could be any constant length as it serves only as a normalization constant. For less than approx. 40 \(L_c\) the parameter \(\kappa\) jumps because the turbulent field on such scales correlates the movement of cosmic rays which tends to create additional patterns within the source image which appear as smaller inner hot spots.}
\end{figure}

Another set of simulations is done to check the applicability of the small angle approximation for a particular physical parameter. Over the distance of 5 Mpc, the forward propagation method is used to propagate 5 EeV energy protons through various realizations of the turbulent magnetic field with \(B_\mathrm{rms} = 5\,\mathrm{nG}\),  \(L_\mathrm{min}= 20\,\mathrm{kpc}\), and \(L_c\) ranging from 28 to 256 kpc. For every \(L_c\) 30 random realizations of the magnetic field are generated to calculate the mean and the variance. Particles are injected from a single source and detected if they hit a small observer sphere from the outside (later referred just as the observer). The radius of the observer is 200 kpc, and it is chosen so to minimize the artificial effect of the finite observer which affects the angular spread but at the same time to register at least a few hundred events (see Appendix \ref{sec:finite-size} for details). Numerically, \(\kappa\) is obtained from the data by using eq. (\ref{eq:kapparho}). Figure \ref{fig:coherencelengths} shows the result from which it can be concluded that at least 40 coherence lengths are needed to reproduce the shape of the spread parameter which corresponds to the analytical form.

\begin{figure}[th]
      \centering
      \includegraphics[scale=0.45]{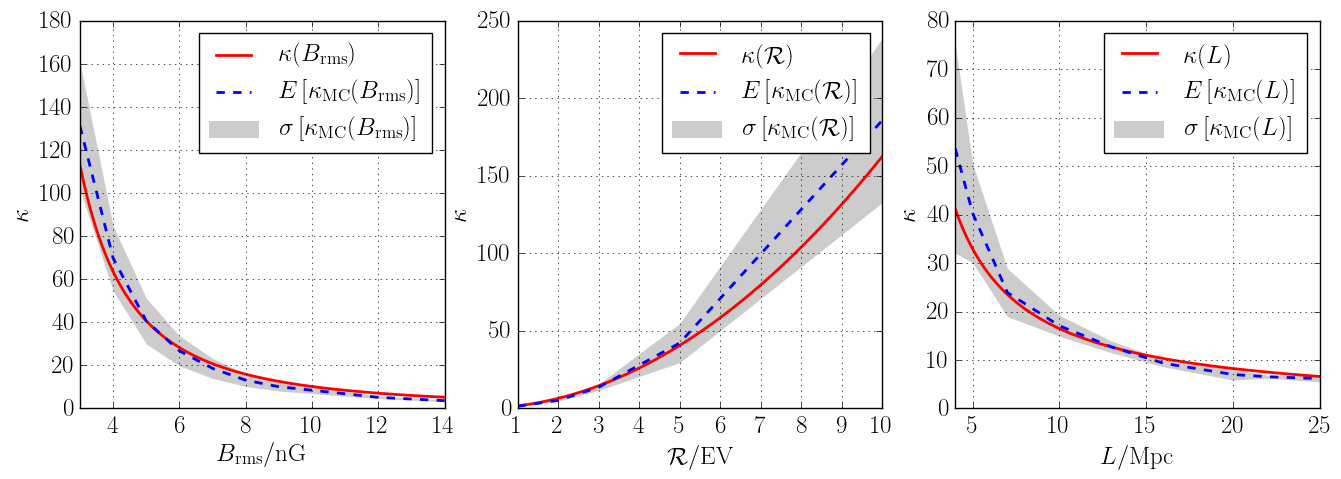}
\caption{The parametrization of equation (\ref{eq:parametrization}) is tested within the CRPropa3 framework \cite{batista2016crpropa} for every parameter magnetic field strength \(B_\mathrm{rms}\), rigidity \(\mathcal{R}\) and distance \(L\) separately. Red lines are calculated from the formula, while blue dashed lines are Monte Carlo results.\label{fig:parametrization}}
\end{figure}

To verify the dependence on the other parameters in equation (\ref{eq:parametrization}) the same approach is used following in part ref. \cite{urbancomparison}. The coherence length of the turbulent field is set to 100 kpc and the other parameters are equated with \(B_\mathrm{rms} = 5\,\mathrm{nG}\), \(\mathcal{R}=5\,\mathrm{EV}\), \(L=5\,\mathrm{Mpc}\) when they are not the subject of the check themselves.

It is worth to note that small discrepancies and fluctuations for \(\kappa \geq 50\) are of less importance because anisotropic moments \(C_\ell\) are not changing much for those values and moderate \(\ell\) which is expected because the large \(\kappa\) means the small deflection to which low multipoles are not very sensitive. That can be argued by inspecting the derivatives \(\dd C_\ell(\kappa) / \dd \kappa\) in the second panel of fig. \ref{fig:vonmisses_plot} (the dashed colour lines) which shows that derivatives tend to 0 as \(\kappa \rightarrow \infty\).

\subsection{The influence of the second nearest source}
\label{sec:second_nearest}

By assuming the dependence of the spread parameter on distance, one can assess the validity of the starting assumption that the distant sources do not sufficiently change the anisotropy created by the closest source. Introducing only two sources in an analogous manner as in equation (\ref{eq:distribution}) with different distances \(L_1\), \(L_2=\lambda L_1\) and with spread parameters \(\kappa_1\), \(\kappa_2=\kappa_1(L_1/L_2)= \kappa_1 / \lambda\) respectively, the calculation from section  \ref{sec:model} can be repeated. The fluxes fall as \(L_{1,2}^{-2}\), hence the relative flux is \(j_2/j_1 = \lambda^{-2}\) if both sources are injecting particles with the same rate. Only the dipole is calculated as the most important moment in any case. Two extreme cases are studied to preserve the axial symmetry which simplify the integration, a ``constructive'' one (\(\hat{\mathbf{r}}_1 = \hat{\mathbf{r}}_2\)), where the second source is behind the first one, and a ``destructive'' one (\(\hat{\mathbf{r}}_1 = -\hat{\mathbf{r}}_2\)), where the second one is on the opposite side, relative to the observer.
\begin{align}
\label{eq:distribution_twosources}
f_\mathrm{src_{1,2}}(\hat{\mathbf{r}}|\lambda, \kappa_1, \kappa_2) =& f_\mathrm{src_{1,2}}(\hat{\mathbf{r}}|\lambda, \kappa_1) = \nonumber \\
&(1-\frac{\lambda^{-2}}{2})
\frac{\kappa_1}{4\pi \sinh(\kappa_1)} \exp(\kappa_1 \hat{\mathbf{r}}\cdot\hat{\mathbf{r}}_1)
\nonumber \\
&+ \frac{\lambda^{-2}}{2}
\frac{\kappa_1/\lambda}{4\pi \sinh(\kappa_1/\lambda)} \exp(\kappa_1 \hat{\mathbf{r}}\cdot\hat{\mathbf{r}}_2 /\lambda) \nonumber \\
\Rightarrow a_{10} =& \sqrt{\frac{3}{4\pi}}\left[
\mathcal{I}_1(\kappa_1) - \frac{1}{2\lambda^2}\left(\mathcal{I}_1(\kappa_1)\pm \mathcal{I}_1(\kappa_1/\lambda) \right)\right]
\end{align}

\begin{figure}[h]
      \centering
      \includegraphics[scale=0.50]{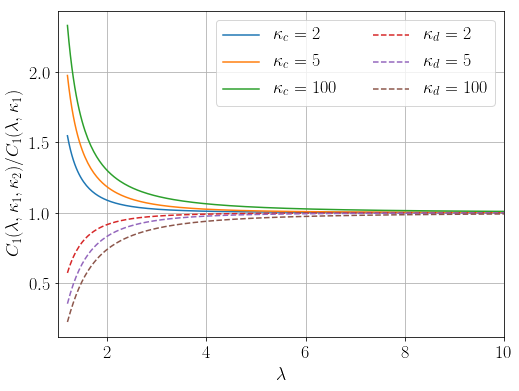}
\caption{The plot shows when the dipole moment converges to the original expression given by (\ref{eq:vonmisses_solution}) if there is a second source involved that contributes to the total flux. Parameter \(\lambda\) is the ratio between the distances of the two sources - the more remote the second source is, the less it interferes with the dipole and other moments created by the first source. Two extreme cases are considered, a ``constructive'' one (solid lines) where the observer, the first source and the second source are located on the same line of sight in that order and a ``destructive'' one (dashed lines) where the observer is located between the two sources.\label{fig:assumption}}
\end{figure}

By requiring \(\lim_{\lambda\rightarrow \infty} C_1(\lambda, \kappa_1,\kappa_2)/C_1(\kappa_1) = 1\), the deviation from the single discrete source case can be parametrised through parameter \(\lambda\). The following solution is plotted in figure \ref{fig:assumption}:
\begin{align}
\label{eq:assumption}
\frac{C_1^\mathrm{src_{1,2}}(\lambda, \kappa_1, \kappa_2)}{C_1^\mathrm{src_1}(\eta=1/2\lambda^2, \kappa_1)} =&
\left[ 1
\pm
\frac{ \coth \left(\frac{\kappa }{\lambda }\right)-\frac{\lambda }{\kappa }}{\left(2\lambda^2-1\right)\left(\coth (\kappa )-\frac{1}{\kappa }\right)}
\right]^2 \nonumber \\
\approx &
1 \pm \frac{1}{3 \lambda^3}\left(3 + \frac{1}{5}\kappa^2 + {O}(\kappa^4)\right)
\end{align}
where the last approximation is valid for \(\lambda \gg 1\) and \(\kappa \sim 1\).

The influence of the second source on the dipole falls with the distance as \(\sim \lambda^{-3}\) for a given spread parameter \(\kappa\), but if the angular spread is larger (smaller \(\kappa\)), the decrease is quicker for both cases. Thus, for the stronger magnetic fields and lower rigidities, the single discrete source assumption holds better, as expected. If the second source is located at two times the distance of the first one, its effect on \(C_1\) is already smaller than 20\%.

\subsection{Combining multiple rigidities}
\label{sec:rigidity}

Up to now, the calculations considered only a monoenergetic source which injects just a single type of cosmic rays, i.e., injects a single rigidity. An expansion of the analysis to multiple rigidities can be represented through a single source which is a combination of multiple sources located at the same position which differ in the spread parameter \(\kappa\). This leads to the following expression of the combined spectrum:
\begin{equation}
\label{eq:almighty}
 C_\ell^\mathrm{comb}(\kappa_1, \dots, \kappa_N) = \frac{1}{4\pi} \left[ \sum_i^N \eta_i \mathcal{I}_\ell (\kappa_i) \right]^2
\end{equation}
where \(\eta_i\) represents the share of the component \(i\) in the total flux and \(\sum_i^N\eta_i = 1\).

Here we state the following claim which validity we argue numerically. If two angular power spectra, the combined one defined with eq. (\ref{eq:almighty}) and the pure one defined with eq. (\ref{eq:vonmisses_solution}), share the same value of the dipolar component \(C_1^\mathrm{comb} = C_1^\mathrm{pure}\), all higher components \(\ell > 1\) of the combined spectrum will have equal or higher values compared to the same component in the pure case.

The first step is to find \(\kappa^\mathrm{pure}\) from a set of \(N\) given pairs \((\eta_i, \kappa_i)\) using \(\mathcal{I}_1(\kappa^\mathrm{pure}) = \sum_i^N \eta_i \mathcal{I}_1(\kappa_i)\) which is derived from \(C_1^\mathrm{comb} = C_1^\mathrm{pure}\). Verifying the above claim is then equal to checking if the following is satisfied:
\begin{equation}
\frac{C_\ell^\mathrm{comb}(\kappa_1, \dots, \kappa_N)}{C_\ell^\mathrm{pure}(\kappa^\mathrm{pure})} \geq 1 \ .
\end{equation}
This is verified in case of the first few multipoles for a wide range of randomly chosen \((\eta_i, \kappa_i)\), and there is no reason why the inequality would not be fulfilled for some other sets since all involved functions are monotonically increasing functions without inflection points (see fig. \ref{fig:vonmisses_plot}). It would be interesting to prove the statement exactly, but this is out of the scope of this work.

\section{Comparison with structured magnetic fields}
\label{sec:montecarlo}

\begin{figure}[h]
      \centering
      \includegraphics[scale=0.45]{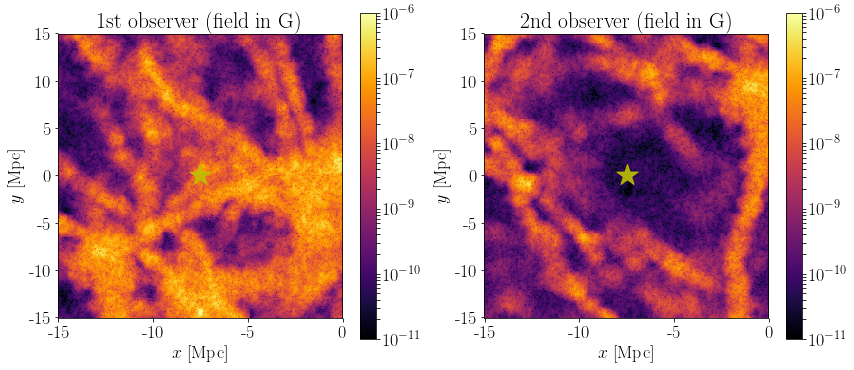}	
\caption{Two distinct locations for the observer are chosen: the first one within structures which is also constrained to resemble the local universe around the Milky Way; and the second one is placed in a void at least 5 Mpc from the nearest structure. The color scale represents the magnetic field strength. The shown magnetic field comes from the benchmark model described in the text. \label{fig:fieldaroundobservers}}
\end{figure}

The above-presented model follows the rather naive assumption that magnetic fields in the local universe are everywhere uniformly turbulent only. Since galaxies, clusters, filaments, and voids contain magnetic fields which differ by several orders of magnitude in strength and other properties, it is more realistic to assume that magnetic fields are structured which is also backed by observations. Nowadays, there are several structured extra-galactic magnetic field models available such as \cite{astro-ph/0412525v1, Dolag:2004kp, Hackstein:2016pwa, AlvesBatista:2017vob}. The magnetic field used here, also described in details in the CRPropa 3 reference paper \cite{batista2016crpropa} and referred later as the benchmark field, is constructed on the basis of work of Dolag et al. \cite{Dolag:2004kp} and Miniati et al. \cite{astro-ph/0412525v1}. It represents a constrained model of the local large-scale structure taken from Dolag et al. on top of which is applied the magnetic field from Miniati et al. by first constructing the correlation density magnetic field within Miniati and then applying it to the Dolag density field. Within uncertainties of the void fraction in the local universe \cite{Colberg:2008qg}, the chosen model's magnetic field strength in voids comply with the newest upper limits, such as from the Planck experiment \cite{Ade:2015cva}, but the more important component to compare here is the structure of the magnetic field itself, not its strength. The magnetic field structure in a form of strength distribution is shown in figure \ref{fig:fieldaroundobservers}. Furthermore, two distinct locations for the observer have been investigated: the first one located within structures which resembles the local universe around the Milky Way; and the second one is placed in a void at least 5 Mpc from the nearest structure (figure \ref{fig:fieldaroundobservers}).

\begin{figure}
\begin{minipage}{0.50\textwidth}
	      \centering
	      \includegraphics[scale=0.30]{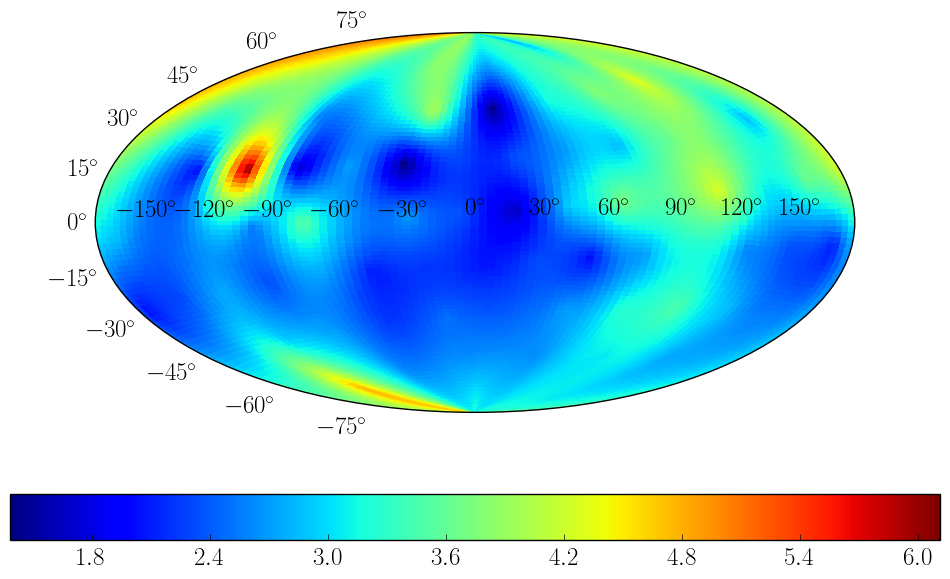}	  
\end{minipage}
\begin{minipage}{0.50\textwidth}
	      \centering
	      \includegraphics[scale=0.30]{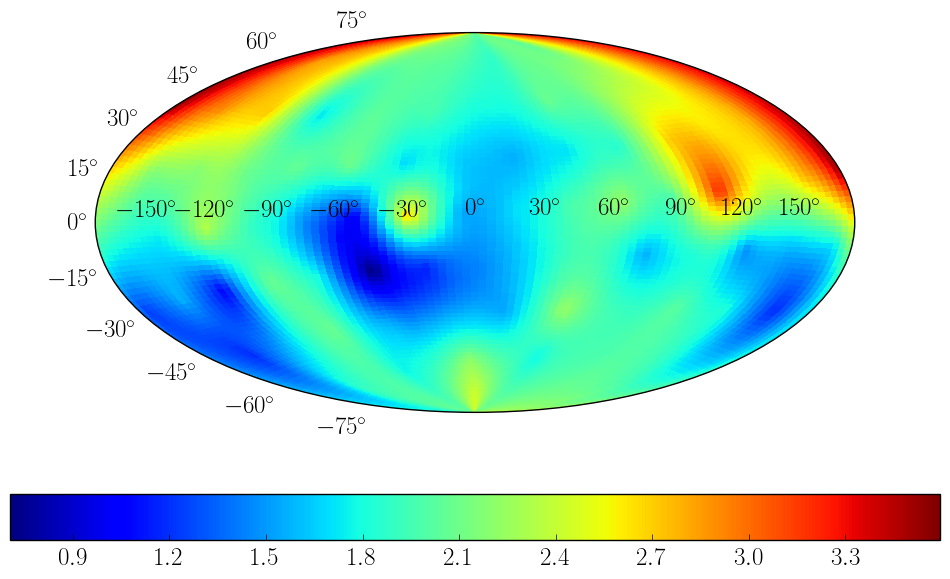}
\end{minipage}
\begin{minipage}{0.50\textwidth}
	      \centering
	      \includegraphics[scale=0.30]{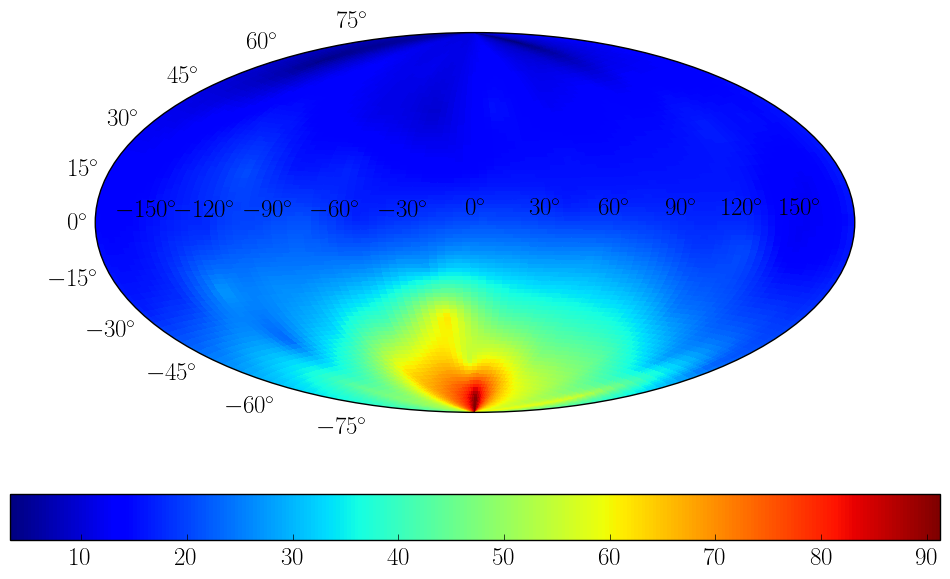}	  
\end{minipage}
\begin{minipage}{0.50\textwidth}
	      \centering
	      \includegraphics[scale=0.30]{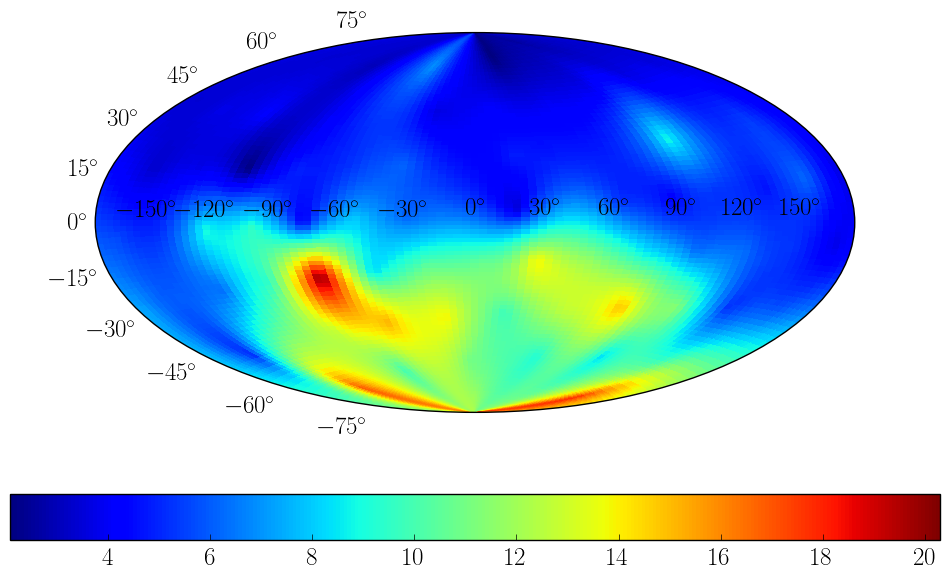}
\end{minipage}
\caption{These four skymaps interpolate values of \(\kappa\) from 80 uniformly distributed sources in four cases (the color scale represents the value of \(\kappa\)): upper two represent skymaps for the first location of the observer (within structures) where sources are at 5 Mpc (left) and 10 Mpc (right), while the bottom two represent skymaps for the second location of the observer (in the void), also at the same distances as the upper two. Average values are \(\kappa_1^{5\,\mathrm{Mpc}} = 2.9 \pm 0.8\), \(\kappa_1^{10\,\mathrm{Mpc}} = 2.0\pm 0.5\), \(\kappa_2^{5\,\mathrm{Mpc}} = 22 \pm 14\) and \(\kappa_2^{10\,\mathrm{Mpc}} = 7 \pm 4\). The angular spread depends on the direction, but generally the same quantitative behaviour is preserved: the stronger magnetic field (the 1st location, the upper row) causes the larger angular spread (smaller \(\kappa\)). The same holds for the greater distance from sources (the right column). The structure of the magnetic field around these locations is displayed in the previous figure (\ref{fig:fieldaroundobservers}). \label{fig:skymaps_kappas}}
\end{figure}

The simulation scenario consists of iterating a single source over 80 uniformly distributed locations on a sphere of radius 5 Mpc and 10 Mpc around the observer. The observer sphere has radius of 400 kpc. The energy range of the particles is chosen such that the spread parameter ranges from \(\kappa \approx 1\) to \(\kappa \approx 10\) which is the range where the effect of the structure in the angular distribution is most easily seen in the context of this investigation, namely around 10 EeV. Other parameters are kept the same as in the turbulent case. At least one thousand events are collected from every source. The results are shown in figure \ref{fig:skymaps_kappas}. The mean spread parameter of all sources in the first location, within structures, is \(\kappa_1^{5\,\mathrm{Mpc}} = 2.9 \pm 0.8\) and \(\kappa_1^{10\,\mathrm{Mpc}} = 2.0\pm 0.5\) for sources at 5 Mpc and 10 Mpc distance, respectively. In the second location, within the void, the values are \(\kappa_2^{5\,\mathrm{Mpc}} = 22 \pm 14\) and \(\kappa_2^{10\,\mathrm{Mpc}} = 7 \pm 4\).

\begin{figure}
\begin{minipage}{0.50\textwidth}
	      \centering
	      \includegraphics[scale=0.30]{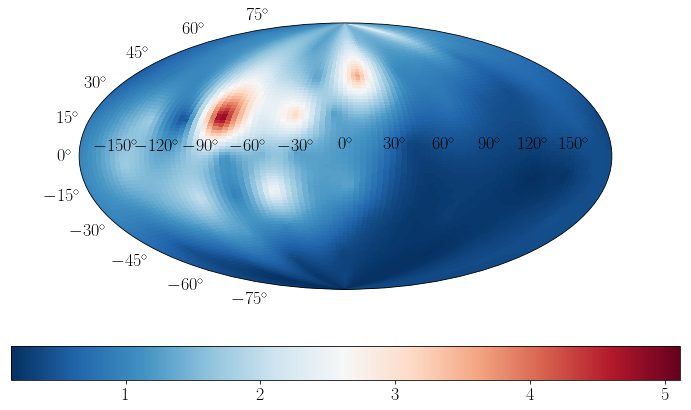}	  
\end{minipage}
\begin{minipage}{0.50\textwidth}
	      \centering
	      \includegraphics[scale=0.30]{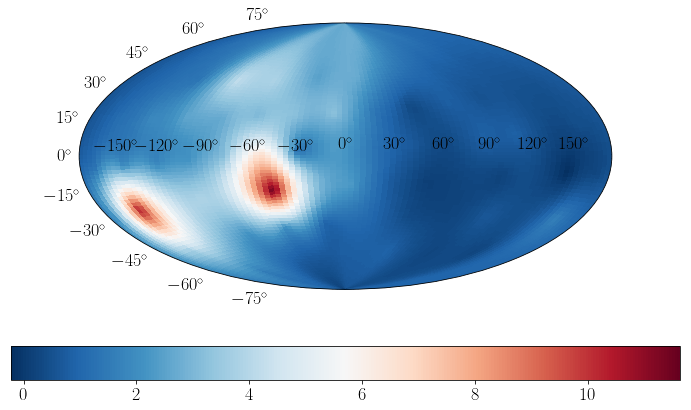}
\end{minipage}
\begin{minipage}{0.50\textwidth}
	      \centering
	      \includegraphics[scale=0.30]{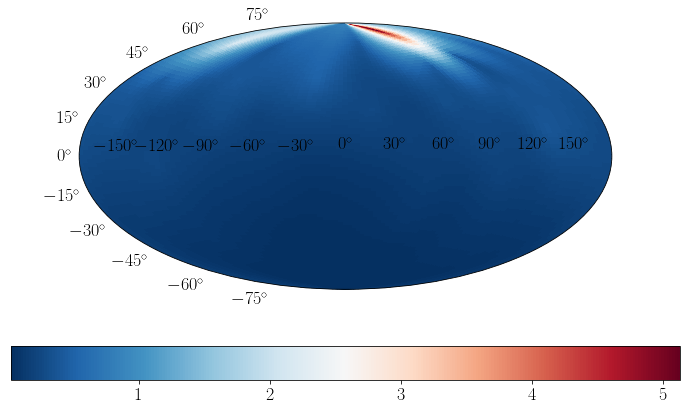}	  
\end{minipage}
\begin{minipage}{0.50\textwidth}
	      \centering
	      \includegraphics[scale=0.30]{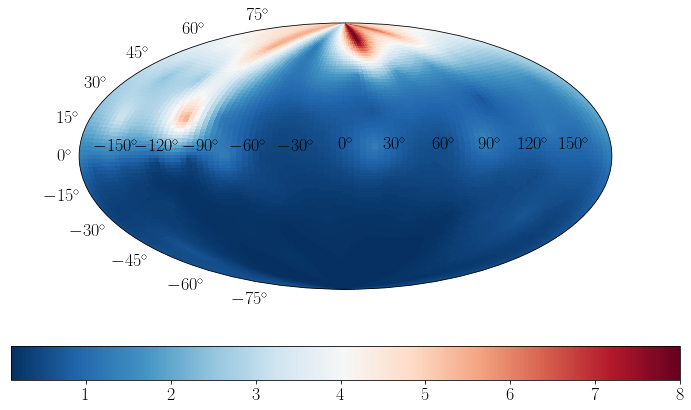}
\end{minipage}
\caption{This figure describes the same cases as the previous one, figure \ref{fig:fieldaroundobservers}, but it shows the relative difference in the quadrupole component between the simulation and the analytical expression, \((C_2^\mathrm{MC} - C_2^\mathrm{AN})/C_2^\mathrm{AN}\) which is represented by the color scale. The smaller the difference is, the better the original pattern (figure \ref{fig:vonmisses_plot}) is reproduced, e.g., the better it resembles the pure turbulent case. Red spots are areas where the structured magnetic field changes the expected ratio between the dipole and quadrupole moment the most. It should be stressed that in all such cases the quadrupole to dipole ratio tends to be larger than predicted for purely turbulent fields.\label{fig:skymaps_quad_diff}}
\end{figure}

In a structured magnetic field it is not possible to unambiguously define physical parameters in the whole space such as \(B_\mathrm{rms}\) which determines the spread parameter as in case of a turbulent field only (equation \ref{eq:parametrization}), but it is possible to investigate how much the shape of the angular power spectrum is changed compared to expected solution \ref{eq:vonmisses_solution}. To this end we compare the next most robust large-scale component, the quadrupole, with its value predicted by eq. (\ref{eq:vonmisses_solution}) for a purely turbulent field producing the same dipolar moment as the structured magnetic field studied here. Taking the relative difference \((C_2^\mathrm{MC} - C_2^\mathrm{AN})/C_2^\mathrm{AN}\) (figure \ref{fig:skymaps_quad_diff}), where marks `AN' and `MC' stands for analytically and numerically obtained values, respectively, it can be observed in which directions the original pattern expected from eq. (\ref{eq:vonmisses_solution}) (figure \ref{fig:vonmisses_plot}) is the least affected by the structures. In the void, the pattern is conserved the best, as expected, but for larger distances where new domains of the magnetic field emerge the shape of the Fisher distribution is disturbed. For the observer in the structured part of the universe, the pattern is mostly distorted on the edges between two different domains of the large and small spread parameters which can be seen by comparing figures \ref{fig:skymaps_kappas} and \ref{fig:skymaps_quad_diff}. In areas where the pattern is distorted, the predicted single source signature is basically lost since different moments are modified differently, yet, it should be stressed that in all such cases the effect of structures increased the quadrupole component compared to the dipole component relative to the analytical expectation. This investigation is by no means complete and certainly not quantitatively stringent, but it sketches what is already intuitively expected.

A qualitative description of the phenomena is now proposed. Generally, a structured magnetic field can be characterized by the existence of separate domains which have different field properties and spatial sizes, therefore, these domains contribute differently to cosmic ray deflection, that is, they have different \(\kappa\). If a source and an observer are located within the same domain, the described pure turbulent field approach can be applied, but if cosmic rays from the source traverse different domains to reach the observer, each domain can be considered as an independent source located at the same position as the real source but with an independent spread parameter. That leads to a power spectrum of the same form as eq. \ref{eq:almighty}. As was already shown in subsection \ref{sec:rigidity} the values of higher multipoles will increase when combining sources of different \(\kappa\) compared to the pure turbulent case. Summarizing the argument, the structured magnetic field tends to enhance higher multipoles of the angular power spectrum compared to the analytically obtained reference spectrum.

In domains where the assumption of traversing multiple coherence lengths is not fulfilled, cosmic rays can create distinct structured patterns like creating mirror images \cite{astro-ph/0202362v2} which will considerably affect the argument. Similar effects would occur if regular components of the magnetic field dominate the turbulent one in those domains. Considering those scenarios is beyond the scope of this work.

\section{Examining constraints from the measured dipole and quadrupole and potential nearby sources}
\label{sec:constraints}

\begin{figure}
\centering
\includegraphics[scale=0.50]{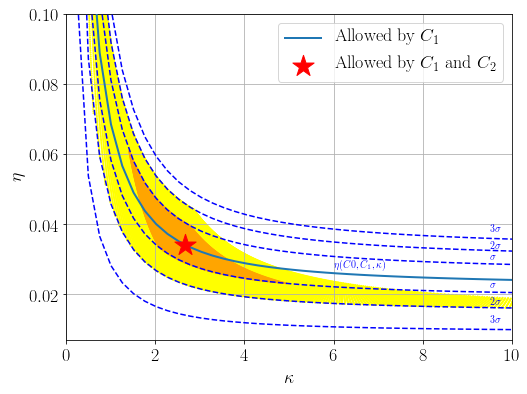}
\caption{The plot shows how the Auger dipole and quadrupole at 8 EeV \cite{Aab:2016ban} constrain the space of parameters \(\eta\) and \(\kappa\). The solid blue line is the result if only the dipole is considered while dashed lines are appropriate uncertainties. If the quadrupole is taken into account, there is one possible solution (given by equations (\ref{eq:mukapa}) and \ref{eq:mukapa2}) marked with the star and corresponding one, and two sigmas shaded areas (orange and yellow respectively). From the plot, it can be concluded that if the anisotropy is caused by a single source a significant angular spread (small \(\kappa\)) is favoured which almost completely erases \(C_\ell\) with \(\ell>2\). The significant angular spread also implies larger deflection for given energies and composition than expected in typical magnetic field models (see equation (\ref{eq:parametrization}) and also the next figure).\label{fig:exclusion}}
\end{figure}

The Pierre Auger Observatory has observed a dipolar amplitude of \(d=6.5^{+1.3}_{-0.9}\%\) above 8 EeV \cite{Aab:2017tyv} while higher moments remain within 99\% confidence level of isotropy \cite{Aab:2016ban}. To derive values of \(C_\ell\) with the correct normalization in the monopole component \(C_0\) for the Auger data the following expression from \cite{Abreu:2012ybu} is used: \(\Phi(\mathbf{\hat{n}}) = \frac{\Phi_0}{4\pi}(1+\mathbf{d}\cdot \mathbf{n})\). Putting it in definition (\ref{eq:ftosharmonics}), one obtains \(C_1/C_0 = |\mathbf{d}|^2/9 = (0.00047^{+0.00019}_{-0.00013})\). As already noted in section \ref{sec:model}, equation (\ref{eq:mukapa}) can serve to constrain the model's parameters from the dipole amplitude. The result is displayed as the solid blue line together with associated uncertainties as the dashed blue lines in figure \ref{fig:exclusion}. Employing the constrained value at 2-\(\sigma\) level of the quadrupole moment from the Pierre Auger experiment with the abovementioned normalization \(C_2/C_0=(1\pm 0.5)\times 10^{-4}\) \cite{Aab:2016ban}, the model gives \(\kappa = 2.7^{+1.2}_{-1.1}\) and \(\eta=0.03\pm 0.01\) where uncertainties are obtained from the experimental uncertainties of \(C_1\) and \(C_2\) by integrating out either \(\eta\) or \(\kappa\). The shaded areas in figure \ref{fig:exclusion} (orange for one sigma and yellow for two sigmas) represent the propagated confidence ranges for every \(\eta\) and \(\kappa\). By inserting \(C_2\) here, we assume that it has only the \(a_{20}\) component which originates from the same direction as the dipolar anisotropy. It should also be stated that the result completely ignores the influence of the galactic magnetic field which could transfer the dipolar anisotropy into the quadrupole and higher multipoles \cite{Giacinti:2011mz}, and, thus, the constrained could be further loosened.

The retrieved \(\kappa\) is larger than 1 which guarantees that the small angle approximation can be used, thereby, the root mean square of the deflection angle \(\alpha_{\rm rms}\) from eq. (\ref{eq:parametrization}) yields:
\begin{equation}
\alpha_{\rm rms} = \left(50^{+11}_{-10}\right)^\circ \quad \left(\mathcal{D}_0 L = 0.56^{+0.25}_{-0.22}\right) \ .
\end{equation}
However, the displayed value of \(\mathcal{D}_0 L\) requires a more beamed source that is injecting cosmic rays towards the observer, not in every direction. Otherwise, the Fisher distribution of the arrival directions assumed at the beginning of this work cannot directly apply here. An isotropic source would for the same dipolar amplitude give smaller \(\mathcal{D}_0 L\) and \(\alpha_{\rm rms}\) (see fig. \ref{fig:comparison_with_harari}).

The large uncertainties are the consequence of the significant derivatives of equation (\ref{eq:mukapa}) in the regime of small \(\kappa\) (see the derivatives in the second plot of figure \ref{fig:vonmisses_plot}) combined with the large experimental uncertainties in the large-scale anisotropy sector which are firstly due to low statistics at the highest energies, and secondly due to limited coherent full-sky coverage \cite{sommers2001cosmic}. In current full-sky analysis \cite{Aab:2014ila} uncertainties in systematics between the two largest experiments, Telescope Array and Pierre Auger Observatory, prevent more precise results than those obtained by reconstructing the large scale anisotropies from partial sky coverage of a single experiment.

The model's parameters derived above imply considerable angular spreads compared to what is usually considered when inserting in eq. (\ref{eq:parametrization}) common estimates of the RMS magnetic field strength, coherence length and distances of neighbouring objects that are hypothesised to be UHECR sources. On the one hand, the significant angular spread reinforce the single source approximation as the influence of the second nearest source falls rapidly with distance (see the equation (\ref{eq:assumption}) and figure \ref{fig:assumption}). On the other hand, the results from section \ref{sec:montecarlo} tend to greatly influence the dipole for the range of the constrained \(\kappa\). Therefore, the ratio \(C_2/C_1\) is increased more when cosmic rays traverse domains of a structured magnetic field for the given spread parameter which consequently impose more stringent conditions on \(\kappa\) and \(\eta\). Or to put it differently, if the measured anisotropies are originating from a single source and the quadrupole component is increased compared to the dipole component due to the structured magnetic field of the real Universe, the consequence is that even smaller \(\kappa\) is required by the observations than the one retrieved here.

By using solution (\ref{eq:vonmisses_solution}) for \(\ell=1\) and parametrisation (\ref{eq:parametrization}) to express the dipole amplitude \(d=\sqrt{9 C_1/C_0}\) as function of rigidity \(\mathcal{R}\), one gets:
\begin{equation}
\label{eq:dipole_comparison}
 d(\mathcal{R}) = 3\eta\left(\coth{\kappa} - \kappa^{-1}\right) \approx 3\eta \left[\coth{\left(\frac{\kappa_0}{\mathcal{R}_0^2}\mathcal{R}^2\right)} - \left(\frac{\kappa_0}{\mathcal{R}_0^2}\mathcal{R}^2\right)^{-1}\right]
\end{equation}
where \(\kappa_0\) and \(\mathcal{R}_0\) are fixed by the dipole measured in the energy bin centered at 11.5 EeV and with an assumption for the averaged composition at that energy of \(\langle Z \rangle = 4\) extracted from \cite{PhysRevD.90.122006}. Thus, the inserted rigidity is \(\mathcal{R}_0 = 2.875\, \mathrm{EV}\) and the spread parameter \(\kappa_0 = 2.7\).

\begin{figure}
\centering
\includegraphics[scale=0.50]{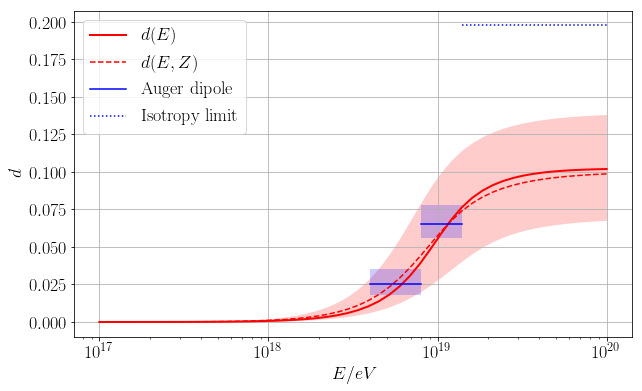}
\caption{\label{fig:auger_dipole_comparison}This is a comparison of the model's result for the dipole with measured dipole amplitudes at two different energy ranges taken from \cite{Aab:2017tyv} by using (\ref{eq:dipole_comparison}). Equation (\ref{eq:dipole_comparison}) with \(\kappa_0\) and \(\eta_0\) determined by the measurement at 11.5 EeV with the assumed composition of \(Z=4\) is plotted as the solid line. The red shaded area represents uncertainties coming from the uncertainties in \(\kappa\). The resulting prediction of the dipole at the lower energy range is consistent with the measured dipole in the 4-8 EeV bin. The dashed line represents a correction in the dipole amplitude \(d(Z)\) when the changing average composition \(\langle Z(E)\rangle\) at higher energies is taken into account. This shows that the measured composition indications slightly suppress the dipole amplitude at the higher energies, while negligibly amplifying it at lower energies. The dotted line is a dipole average of the isotropic distribution in the case of full-sky coverage when roughly the total number of events detected by PAO, \(N\approx 200\), above \(10^{19.5}\,\mathrm{eV}\) \cite{PhysRevD.90.122006} is inserted in \(d = \sqrt{9 \langle C_1\rangle_{\rm iso} / C_0} = 3 N^{-1/2}\) \cite{sommers2001cosmic}.}
\end{figure}

An immediate consequence is that the dipole amplitude should increase at higher energies. The plot of the function and the Auger dipolar amplitudes for 4-8 EeV and 8+ EeV bins can be seen on fig. \ref{fig:auger_dipole_comparison}. The measured dipole amplitude in the 4-8 EeV range is in agreement with the derived result. The composition data from the Pierre Auger Observatory \cite{PhysRevD.90.122006} suggests a change in the average composition at the end of the spectrum towards heavier nuclei when the fit of the EPOS-LHC hadronic model to the shower data is used. Here, just to illustrate the change in the composition, the composition dependence on energy is simplified to roughly match the one of the EPOS-LHC hadronic model in \cite{PhysRevD.90.122006}. The composition is averaged per energy bin \(\langle Z\rangle = \sum_{i=\mathrm{p}, \mathrm{He}, \mathrm{N}, \mathrm{Fe}} f_i Z_i\), and fitted to a linear function resulting with \(\langle Z(E)\rangle = 4 + 0.1 E/\mathrm{EeV}\). When this dependence is inserted in \(d(\mathcal{R})\), the dipole amplitude at higher energy bins is slightly suppressed compared to the prediction without this dependence. Other hadronic models from \cite{PhysRevD.90.122006} reduce this difference since they predict a more uniform composition in this energy range, from \(10^{17}\) to \(10^{20}\) EeV. Uncertainties in the composition estimate are not taken into account due to the large uncertainties in the spread parameter which dominate in the dipole prediction in any case (the red shaded area on the figure).

In the field of ultra-high energy cosmic rays, two nearby extra-galactic objects often considered as potential sources are Centaurus A \cite{cavallo1978sources} and Virgo cluster \cite{mathews2011cosmic, Dolag:2008py}. This conclusion comes from reasoning that if our galaxy does not contain known accelerators which are capable of achieving energies above 8 EeV \cite{cavallo1978sources}, it follows that ordinary galaxies cannot produce UHECRs either. Therefore, radio galaxies or rich clusters of galaxies are more likely to contain UHECR sources. The radio galaxy Centaurus A and Virgo cluster are the two closest of these kinds. The distance from Cen A is \(3.8\ \mathrm{Mpc}\) \cite{harris2010distance} and from Virgo \(16.5\ \mathrm{Mpc}\) is \cite{mei2007acs}.

\begin{figure}
  \centering
  \includegraphics[scale=0.50]{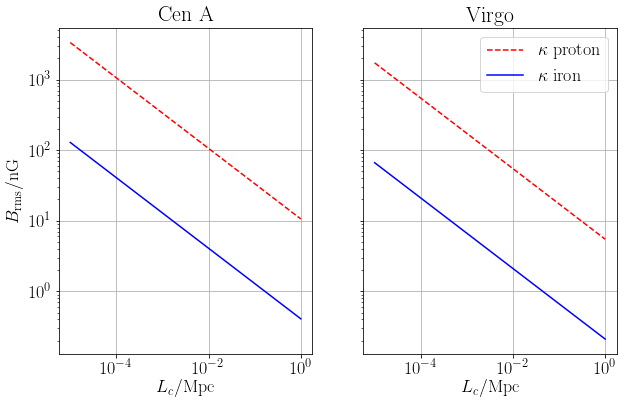}
  \caption{For a source of a given distance, the remaining parameters left undetermined are charge, magnetic field strength and coherence length. The plot shows the relation between \(B_{\rm rms}\) and \(L_c\) following from eq. (\ref{eq:parametrization}) for the fitted value of \(\kappa\), for proton and iron primaries coming from Centaurus A and the Virgo cluster.\label{fig:constrains}}
\end{figure}

Putting their distances into eq. (\ref{eq:parametrization}) and using the constraints from the previous paragraph, the quantity \(Z^{-2} B_\mathrm{rms}^{-2}(L_c)^{-1}\) is constrained. This can be seen in the figure \ref{fig:constrains} where two limiting cases, hydrogen and iron, are shown.

\section{Conclusions}
\label{sec:conclusions}

In the last decade, the theoretical research field of ultra-high energy cosmic rays is predominantly conducted by Monte Carlo simulations. This approach is motivated by the tremendous number of involved processes and effects combined with many poorly known parameters. The distribution of arrival directions, as one of the most important observables in the field, is especially sensitive to all of these uncertainties. At the same time, experiments observe a high level of isotropy in the arrival directions. Since most of the extra-galactic source models predict isolated sources in the neighbourhood of our Galaxy, the observed isotropy could be an indication of strong deflections in the intervening magnetic fields. Potentially significant magnetic field strengths combined with the heavier composition above \(10^{19} \mathrm{eV}\) make the whole scenery even more complicated since it influences the propagation greatly while the ability to determine those fields experimentally is still limited. Consequently, ``hot spots'' originating from sources could be smeared out considerably by the magnetic fields. The smearing dilutes those small-scale anisotropic patterns, which are typically expected from sources nearby, and also contributes to the large-scale anisotropies. In particular, the dipolar anisotropy detected by the Pierre Auger Observatory, as the only statistically significant anisotropy detected so far, could be caused by a source nearby while the higher moments could be influenced by the magnetic fields. In this paper, we test this possibility with analytical tools supported by Monte Carlo simulations.

A simplified model of one source nearby normalized with the isotropic background is studied. The model is based on a cautious derivation of the arrival direction distribution starting from a random walk in flight direction of a cosmic ray. For large deflections in arrival directions, to keep the analytical transparency of the results, the single source is restricted to a beamed source directed towards an observer. Focusing on the large-scale anisotropies as the most robust patterns and without specifying the exact strength of the turbulent field or rigidity of cosmic rays, we investigate the validity domain of the model and draw several conclusions.

It is shown that the scenario predicts the existence of higher multipole moments besides the dipole where the ratio of subsequent moments is determined by the spread parameter. Moreover, the absolute values of the dipolar and the quadrupolar amplitudes are sufficient to determine the model's parameters completely. Using that fact and the recent experimental data from the Pierre Auger experiment, constraints on the parameters can be placed (fig. \ref{fig:exclusion}), although with significant uncertainties because the number of detected events in the given energy range is not big enough to determine a statistically significant quadrupole moment. Nevertheless, even the current data require a considerable angular spread of a single source nearby. The fit of the model to the data gives the following solution for the spread parameter \(\kappa = 2.7^{+1.2}_{-1.1}\) (\(\alpha_{\rm rms} = \left(50^{+11}_{-10}\right)^\circ\)) with the relative flux from the single source \(\eta=0.03\pm 0.01\). The obtained result implies considerable deflections which affect other, more distant, sources as well. If their distance is just \(\sim 2\) times the distance of the first source from the observer, they influence the first source dipolar component by less than 10\% and higher multipoles even less, thus, they contribute to the arrival directions almost isotropically and most of the memory about their position is practically lost (see eq. \ref{eq:assumption} and fig. \ref{fig:assumption}). In contrast, sources at similar distances from the observer invalidate the starting assumption of a single isolated source, and that would require a different kind of analyses. The consistency of the analysis is confirmed in a study of the dipolar amplitude dependence on energy (eq. \ref{eq:dipole_comparison}) which agrees with the measured dipolar amplitudes at other energies, i.e., in the 4-8 EeV bin.

We also applied the constraints derived above from energies above \(8\,\mathrm{EeV}\) to properties of the turbulent extra-galactic magnetic field (eq. \ref{eq:parametrization}). This suggests that if the magnetic field is weaker than \(1\,\mathrm{nG}\) in case of iron nuclei, or \(10\,\mathrm{nG}\) in case of protons, for the coherence length of the field \(\le 1\,\mathrm{Mpc}\), it is likely that there is no single luminous source in the vicinity of the Milky Way, such as Centaurus A or the Virgo cluster.

The comparison between the analytical model predictions and the predictions of Monte Carlo simulations in a structured magnetic field (sec. \ref{sec:montecarlo}), as a more realistic scenario of the real Universe, shows that structures generally do not change the presented analytical pattern significantly, and in those cases when they do, they amplify higher multipoles in the angular power spectrum making the observed dipole above 8 EeV and the quadrupole upper limit even more constraining for the model. The same applies if the source injects mixed composition in a wide range of energies.

\vspace{0.5cm}
\textit{Acknowledgments}. A.D. would like to thank the following people for useful conceptual discussions: Marcus Wirtz for the dipole and related subjects, Stefan Hackstein for extragalactic magnetic fields, and David Wittkowski for Monte Carlo UHECR simulations. We would also acknowledge the referee for carefully and critically reviewing the manuscript and pointing to the weaknesses in its earlier versions. This work was supported by the Bundesministerium für Bildung und Forschung (BMBF).

\begin{appendix}

\section{The finite-size observer problem\label{sec:finite-size}}

\begin{wrapfigure}{r}{7.5cm}
\centering
\begin{tikzpicture}
\draw (0,0) node[circle,fill,inner sep=1pt,label=below:$O$](obs) {};
\draw (4,0) node[circle,fill,inner sep=1pt,label=below:$S$](src) {};
\draw (obs) node[circle,draw,minimum size=2cm](sph) {};

\draw[->] (src) -- node[below] {d} (obs);
\coordinate (tan) at (tangent cs:node=sph,point={(src)},solution=1);
\draw (obs) -- node[left] {r} (tan);
\draw[red,->] (src) -- (tan);
\draw[green,<-] (obs) -- ($(src)-(tan)$);
\draw ($(src)-(tan)$) node[circle,fill,inner sep=1pt,label=below:$S'$](newsrc) {};

\begin{scope}
  \path[clip] (src) -- (tan) -- (obs) -- cycle;
  \draw[red, label=left:\textcolor{blue}{$D$}] (src) circle (1.5cm);
  \coordinate [label={[label distance=1cm, yshift=0.15cm]left:{$\theta$}}] (theta) at (src);
\end{scope}

\end{tikzpicture}
\caption{\label{fig:finite-size} The particle that arrived from $S$ and hit the edge of the observer $O$ (the red vector) is detected as a particle that came from source $S'$ from the perspective of a point-like observer (the green vector). It can be said that the finite-size observer $O$ creates a virtual source $S'$, which is displaced at most by the radius $r$ of observer's sphere. Hence, the finite-size observer transforms a point-like source $S$ into a circular area around $S$.}
\end{wrapfigure}
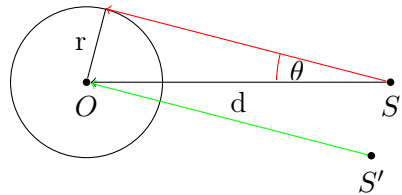

In the case of 3D forward tracking simulations of individual particle trajectories, rescaling the size of an observer is required to match today's computer capability with a decent number of detections. If the observer would be the size of the Earth as in the real Universe, an injection of \(N_{\rm inject} = N_{\rm Earth} (d_{\rm src}/r_{\rm Earth})^2 \sim ({\rm Mpc}/10^6{\rm m})^2 = 10^{33}\) particles would be necessary to have just a few detections from a nearby source which is a megaparsec away. This is unrealistic to simulate since CRPropa3 optimally, without any interactions, can process \(\sim 10^5\,{\rm particles}/{\rm s}/{\rm thread}\) on modern hardware. Same applies if one models the observer as big as the size of bigger objects, like the Galaxy, ~30 kpc in radius, to take into account the observer's volume covered due to the motion of the Earth, the solar system, etc. during the observation period.

However, enlarging the observer to increase statistics introduces unwanted artifacts, especially when anisotropies are studied. From geometrical considerations alone, without deflections, it can be seen (fig. \ref{fig:finite-size}) that a point-like source becomes a disk shape of the arrival directions which artificially changes anisotropy. The maximum angle of the artificial deflection is given by \(\theta_\mathrm{max} = \arcsin\left(\frac{r}{d}\right)\) where \(r\) is the observer radius, and \(d\) is the distance from the source.

To compute the angular power spectrum of a point source seen by a spherical observer we approximate the source image as a homogeneous circle, although it is not perfectly homogeneous but has a radial distribution. A spherical cap is then a suitable model for the case: \(f(\vartheta,\varphi) = \Theta(\theta - \vartheta)\) where $\Theta$ is the Heaviside step function and $\theta$ is the angular size of the circle. The $C_\ell$ can be calculated:

$$a_{\ell m} = \int_0^{2\pi}d\varphi\int_0^\pi \,d\vartheta\,\sin\vartheta \Theta(\theta -\vartheta ) Y_\ell^{m*} (\vartheta,\varphi)$$
$$a_{\ell m} = \int_0^{2\pi}d\varphi\int_0^\theta \,d\vartheta\,\sin\vartheta Y_\ell^{m*} (\vartheta,\varphi)
= 2\pi \sqrt{\frac{2\ell+1}{4\pi}} \int_{\cos(\theta)}^1 \,dx\,P_\ell^0(x)$$

$$ \int_{\cos(\theta)}^1 \,dx\,P_\ell^0(x) = \int_{\cos(\theta)}^1 \,dx\,P_\ell(x) = 2^\ell\cdot \sum_{k=0}^\ell {\ell \choose k}{\frac{\ell+k-1}2\choose \ell} \int_{\cos(\theta)}^1 x^k  \,dx $$
$$ = 2^\ell\cdot \sum_{k=0}^\ell {\ell \choose k}{\frac{\ell+k-1}2\choose \ell} \frac{1}{k+1} \left( 1 - \cos(\theta)^{k+1} \right)$$
$$\Rightarrow C_\ell =   2^{2\ell} \pi \left[ \sum_{k=0}^\ell {\ell \choose k}{\frac{\ell+k-1}2\choose \ell} \frac{1}{k+1} \left( 1 - \cos(\theta)^{k+1} \right) \right]^2 $$

\begin{figure}
  \centering
  \includegraphics[scale=0.50]{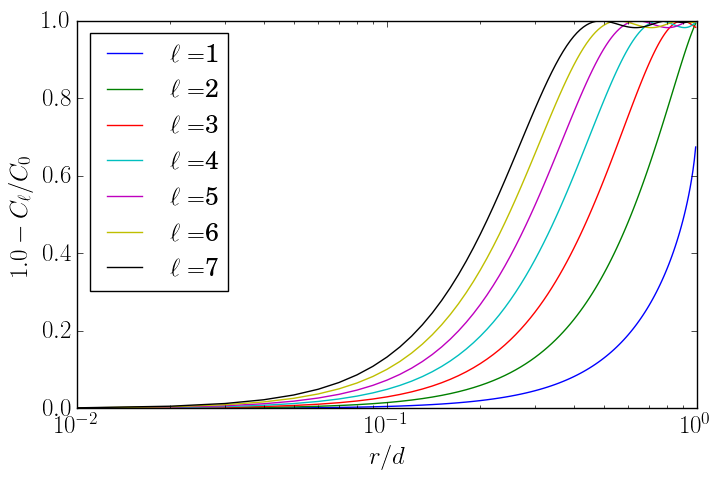}
  \caption{The plot shows how the finite size observer or radius \(r\) changes the angular power spectrum of a nearby source at the distance \(d\). The spectrum without deflections, i.e., when arrival directions have a \(\delta-\)distribution, should be \(C_\ell/C_0 = 1\) without deflections (where \(C_0\) is the monopole component), but as the ratio \(r/d\) grows to unity multiple moments, starting from higher ones, reduce their values in the spectrum. The ratio \(r/d\sim 0.1\) will not influence the dipole, but will deform \(\ell = 7\) by more than 10\%.\label{fig:finite-size-cl}}
\end{figure}

Inserting $\theta_\mathrm{max} = \arcsin\left(\frac{r}{d}\right)$ into \(C_\ell\) and using $\cos(\arccos(x))=\sqrt{1-x^2}$ gives:
\begin{equation}
 C_\ell =   2^{2\ell} \pi \left[ \sum_{k=0}^\ell {\ell \choose k}{\frac{\ell+k-1}2\choose \ell} \frac{1}{k+1} \left( 1 - \left(1-\frac{r^2}{d^2}\right)^{(k+1)/2} \right) \right]^2\ .
\end{equation}
which is plotted in fig. \ref{fig:finite-size-cl} for the first lowest multipole moments. Every moment is expressed relative to the monopole \(C_0 = \pi\left(1-\sqrt{1-\frac{r^2}{d^2}}\right)^2\). To study multipoles below \(\ell = 5\) while keeping the artificial angular spread below 10\% the observer size would always have to be below 10\% of the distance to the nearest source.

\end{appendix}

\bibliographystyle{JHEP}
\bibliography{bibliography}

\end{document}